\documentclass[%
 aip,
 jmp,%
 amsmath,amssymb,
 reprint,%
]{revtex4-1}

\usepackage{graphicx}
\usepackage{caption}
\usepackage{subcaption}
\usepackage{dcolumn}
\usepackage{bm}
\usepackage[usenames,dvipsnames]{color}
\usepackage{amssymb}
\usepackage{verbatim}
\usepackage{hyperref}
\usepackage{listings}
\usepackage{units}
\usepackage{fancyhdr}
\usepackage{csquotes}
\newcommand*\obar[2][0.75]{
    \sbox{\myboxA}{$\m@th#2$}%
    \setbox\myboxB\null
    \ht\myboxB=\ht\myboxA%
    \dp\myboxB=\dp\myboxA%
    \wd\myboxB=#1\wd\myboxA
    \sbox\myboxB{$\m@th\overline{\copy\myboxB}$}
    \setlength\mylenA{\the\wd\myboxA}
    \addtolength\mylenA{-\the\wd\myboxB}%
    \ifdim\wd\myboxB<\wd\myboxA%
       \rlap{\hskip 0.5\mylenA\usebox\myboxB}{\usebox\myboxA}%
    \else
        \hskip -0.5\mylenA\rlap{\usebox\myboxA}{\hskip 0.5\mylenA\usebox\myboxB}%
    \fi}
\makeatother

\renewcommand{\vec}[1]{\boldsymbol{#1}}

\newcommand{\p}{\ensuremath{\partial}}

\renewcommand{\d}{\ensuremath{\mathrm{d}}}

\newcommand{\perfect}{\textsc{Perfect}}

\usepackage{stackengine}
\stackMath

\begin{document}

\preprint{AIP/123-QED}

\title[Neoclassical transport in pedestal with impurities]{Global effects on neoclassical transport in
  the pedestal with impurities }

\author{I. Pusztai}
 \email{pusztai@chalmers.se} 
\author{S. Buller}%
\affiliation{Department of Physics, Chalmers University of Technology,
  SE-41296 G\"{o}teborg, Sweden} 
\author{M. Landreman}
\affiliation{Institute for Research in Electronics and  Applied Physics,
  University of Maryland, College Park, MD 20742, USA.  }

\date{\today}

\begin{abstract}
We present a numerical study of collisional transport in a tokamak
pedestal in the presence of non-trace impurities, using the radially
global $\delta f$ neoclassical solver \perfect{} [M. Landreman et
  al. 2014 \emph{Plasma~Phys.~Control.~Fusion {\bf 56}} 045005]. It is
known that in a tokamak core with non-trace impurities present the
radial impurity flux opposes the bulk ion flux to provide an ambipolar
particle transport, with the electron transport being negligibly
small. However, in a sharp density pedestal with sub-sonic ion flows
the electron transport can be comparable to the ion and impurity
flows. Furthermore, the neoclassical particle transport is not
intrinsically ambipolar, and the non-ambipolarity of the fluxes
extends outside the pedestal region by the radial coupling of the
perturbations. The neoclassical momentum transport, which is finite in
the presence of ion orbit-width scale profile variations, is
significantly enhanced when impurities are present in non-trace
quantities, even if the total parallel mass flow is dominated by the
bulk ions.

\end{abstract}

\pacs{ }
\maketitle


\section{\label{sec:intro} Introduction}
The global confinement in tokamaks is strongly correlated with the
performance indicators of the edge transport
barrier \cite{hubbard00,groebner06}. Accordingly, having a good
pedestal performance in a fusion reactor is considered to be crucial.
In the pedestal, turbulent transport is dramatically reduced by
decorrelation of turbulence by strong sheared
flows \cite{terry00,highcock12}, thus the relative role of the
(otherwise negligible) collisional transport becomes more important.

Fusion reactors will need to operate with carbon-free and low-erosion
plasma facing components to achieve reasonable lifetimes and for
nuclear safety. To gain more operational experience in the proposed
metallic wall devices, originally carbon walled tokamak experiments,
such as JET and ASDEX have transitioned to operation with ``ITER-like
walls'' (with Be and W components) in recent years. With this
transition a global confinement degradation has been
observed \cite{JETAUGILWConfinement2013,JETILWconfinement2014},
especially on JET. This is a serious concern, since the data on which
experimental scaling relations (and projections for ITER) are based
have been mostly collected on carbon wall experiments. This degradation
is found to be primarily caused by reduced pedestal
performance \cite{JETAUGILWConfinement2013,JETILWconfinement2014}.
Impurity injection appears to have a beneficial effect in certain
cases \cite{0029-5515-55-2-023007}. Over and above the general
importance of collisional impurity transport, these observations
motivate consideration of the effects of non-trace impurities in the
pedestal.

Modeling of neoclassical transport in the pedestal is difficult,
since the very sharp profile variations make gradient scale lengths
comparable to the radial extent of the ion drift orbits \cite{kagan09Thesis},
which renders the usual radially local modeling inadequate
for the problem. It has been demonstrated analytically that finite
orbit width effects can significantly modify neoclassical transport
phenomena \cite{kagan08,kagan10PPCF,kagan10,pusztai10,catto11,catto13},
producing modified ion heat transport, flows, bootstrap current, and
even zonal flow response. These studies are limited to large
aspect-ratio, circular cross-section plasmas and describe profile
variations mostly through local plasma parameters, and point at the
importance of a short scale global variation of flows \cite{catto13}.
In this paper we go beyond the large aspect-ratio limit, and present a
radially global numerical study of neoclassical transport in the
presence of non-trace impurities.

There are different possible modeling options of various
sophistication and difficulty.  The simplest option is the
computationally cheap and usually adopted \emph{local $\delta f$}
formalism, which assumes small orbit width compared to profile length
scales and neglects the radial coupling of the perturbations.  On the
other end, the \emph{global full-$f$} approach includes both
neoclassical and turbulent transport (these transport channels cannot
be completely decoupled in a pedestal, in contrast to the local limit)
and allows for strong deviations from local thermodynamic
equilibrium. This approach requires nonlinear collision operators to
fully live up to its promises\cite{candy11}, and is currently
numerically too expensive to be used for exploratory
studies\cite{koh12,seo14}.  In this work, we use the \emph{global
  $\delta f$} formalism \cite{landreman2014}. This is a specific
generalization of the local $\delta f$ formalism which includes global
effects, while still allowing the distribution function to be
linearized around an appropriately chosen lowest order
Maxwellian. This linearization, which assumes sub-sonic parallel
flows, imposes limitations on the profiles that can be
considered. Accordingly, we restrict ourselves to using suitable model
profiles, but with experimentally identifiable features.

In core plasmas the neoclassical perturbations of the ion distribution
are only weakly affected by the electrons. We find that neoclassical
particle transport in a pedestal with subsonic ion flows can be very
different from that in the plasma core (\autoref{fig:fluxon}a-c): The
electron particle flux can be comparable to the ion particle flux even
in the presence of non-trace impurities, and the collisional particle
transport is not ambipolar in general. Consequently, it can happen
that ions and impurities are transported in the same direction. In the
presence of sharp profile variations the neoclassical momentum
transport is nonzero, and we observe that its magnitude is notably
affected by non-trace impurities (\autoref{fig:conc_momsum}a).

The remainder of this paper is organized as follows. In
\autoref{sec:method}, we describe the global $\delta f$ method
implemented in \perfect{}, and how this affects our choice of model
profiles. In \autoref{sec:simulations} we first present the
neoclassical fluxes, flows and poloidal density variations for our
baseline case, and then compare the results between plasmas with trace
and non-trace impurity content. Finally, in \autoref{sec:discussion}
we discuss our results and conclude.


\section{\label{sec:method} Methods}

In a tokamak core the plasma parameter profiles often exhibit
sufficiently slow radial variations that the departure of the
distribution function from a Maxwellian remains small, and the
collisional dynamics can be described in terms of local plasma
parameters. In this situation neoclassical transport can be calculated
using the local $\delta f$ formalism that yields a linear system for
the perturbed distribution. If the driving radial gradients are strong
enough to generate substantial parallel particle flows, and poloidal
variation in plasma parameters, the problem becomes nonlinear
\cite{fulop01}. Furthermore, if all plasma parameters are allowed to
vary over a radial width of a typical ion orbit, the transport becomes
radially non-local. To study such general situations a global full-$f$
simulation code with a nonlinear collision operator would be
necessary. While a limited number of such simulation codes exist
\cite{xu08,koh12,dorf13}, their computational expense make them unfit
for our exploratory purposes. To keep the problem tractable, we will
only consider situations when the distribution functions are not far
from Maxwell-Boltzmann distributions, but finite orbit width effects
are still important.  For this purpose, we use the radially global,
$\delta f$, Eulerian neoclassical solver \perfect.

The fact that the distribution functions must be close to Maxwellians
puts constraints on the profiles, as will be discussed in the
following section. These constraints might not typically be satisfied
in an experiment.  Therefore, we do not attempt to base our
exploratory modeling on specific experimental profiles, instead we use
model profiles chosen specifically to satisfy the assumptions
in \perfect, while they are supposed to be representative of
experimental profiles in some respects. The specific profiles we use
are presented in Appendix~\ref{sec:profilesapp}.  To explain the
origin of the constraints, the next section contains a brief summary
of the equations solved in \perfect{} (for a more detailed description
of the code, we refer the reader to \citenum{landreman2014}).

In addition to the constraints outlined below, \perfect{} does not
capture the geometry of an X-point or the open field line
region. Orbit losses \cite{dorf15APS} and an influx of neutral
atoms \cite{fulop02} are expected to become important very close to
the separatrix. For this reason we expect that our results are
representative only of the inner part of the pedestal.


\subsection{\label{ssec:briefPerfect} The global $\delta f$ problem 
solved by PERFECT }

\perfect{} solves for the non-adiabatic perturbed distribution function
\begin{equation}
  g_a = f_a - f_{Ma} + \frac{e_a \Phi_1}{T_a} f_{Ma},\label{eq:g_a}
\end{equation}
where $f_a$ is the distribution function, $e_a$ is the charge, and
$T_a$ is the temperature of species $a$, $\Phi_1=\Phi-\Phi_0$ is the
perturbed potential, with the unperturbed electrostatic potential
$\Phi_0$ taken to be a flux function,
$\Phi_0=\langle \Phi \rangle$. The flux surface average is defined as
$\langle X \rangle = \int_0^{2\pi}Xd\theta
(\vec{B}\cdot \nabla \theta)^{-1} / \int_0^{2\pi}d\theta
(\vec{B}\cdot \nabla \theta)^{-1}$, where $\theta$ is a
$2\pi$-periodic angle-like poloidal coordinate and $\vec{B}$ is the
magnetic field.  The perturbation $g_a$ is required to be small
compared to the lowest order distribution function, which is a
Maxwell-Boltzmann distribution
\begin{equation}
  f_{Ma} (\psi,W_{a0})= \eta_a(\psi) \left(\frac{m_a}{2\pi
  T_a(\psi)}\right)^{3/2} e^{-\frac{m_a W_{a0}}{T_a(\psi)}},\label{eq:fma}
\end{equation}
where the radial coordinate $\psi$ is $1/(2\pi)$ times the poloidal
magnetic flux, $m_a$ is the mass, $m_{a} W_{a0} = m_av^2/2 + e_a
\Phi_0$ is the total unperturbed energy, and $\eta_a = n_a(\psi)
e^{e_a \Phi_0(\psi)/T_a(\psi)}$ is the pseudo-density, with the
density $n_a$.  The linearized equation that \perfect{} solves is
\begin{align}
  \left(v_\parallel \vec{b} + \vec{v}_{da0}\right)\cdot\left(\nabla
g_a\right) -& C_{la}(g_a) - S_a \nonumber \\ =
-& \vec{v}_{ma} \cdot \nabla \psi \left(\frac{\p
f_{Ma}}{\p \psi}\right),\cite{landreman2014}\label{eq:dke}
\end{align}
where $\vec{b}= \vec{B}/|\vec{B}|$, $v_\|=\vec{v}\cdot \vec{b}$ with
the velocity $\vec{v}$, the lowest order drift velocity
$\vec{v}_{da0}$ contains the lowest order $E\times B$ drift and
magnetic drifts $\vec{v}_{ma}$, $C_{la}$ is the linearized
Fokker-Planck operator, and $S_a$ is a source term, which will be
explained shortly.  The partial derivatives are taken at fixed
magnetic moment $\mu_a=m_a v^2/(2B)$ and unperturbed total energy
$W_{a0}$.

Note that boundary conditions in $\psi$ are needed to fully specify
$g_a$ by \eqref{eq:dke}, in contrast to the local equation. Since the
local theory should apply sufficiently far from the pedestal, the
result of local simulations -- in which the
$\vec{v}_{da0} \cdot \left(\nabla g_a\right)_{\mu,W_{a0}}$ term is
dropped from \eqref{eq:dke} -- are imposed as boundary conditions
where particles enter the domain \cite{landreman2014}.

As inputs, \perfect{} requires zeroth order (flux function) densities
$n_a$, temperature $T_a$ and potential $\Phi_0$. Given these
equilibrium profiles, $g_a$ is calculated from \eqref{eq:dke}, and
appropriate velocity moments of $g_a$ provide the neoclassical flows
and fluxes. The fluxes will in general not be divergence free and thus
incompatible with the time-independent equilibrium profiles. It may
seem instructive to restore the time derivative in \eqref{eq:dke}, and
solve a time-dependent problem, in a hope to reach a steady state
equilibrium. However, only exceptional profiles would lead to a steady
state solution. More generally, particles and energy would accumulate
in (or leave) the simulation domain until the $\delta f$ approach
breaks down. Instead, the approach adopted is to add spatially varying
sources $S_a$ so that the zeroth-order profiles become consistent;
these sources are solved for in the code alongside $g_a$. These
sources can be thought of as representing the effects of
non-neoclassical transport needed to make the profiles consistent, and
should also be present in a real pedestal.

To guarantee that $g_a \ll f_a$, the driving gradients in the
right-hand side of \eqref{eq:dke} should remain small. From
\begin{equation}
   \left.\frac{\p f_{Ma}}{\p \psi}\right|_{W_{a0}}
   = \left[\frac{\eta_a'}{\eta_a} + \left(\frac{m_a W_{a0}}{T_a}
   - \frac{3}{2}\right)\frac{T_a'}{T_a}\right]f_{Ma},\label{eq:dpsifma}
\end{equation}
where prime denotes the $\psi$-derivative, we see that the
$\eta$ and temperature gradients set the size of $g_a$, and thus
drive the deviations from a Maxwellian. Hence the density and the
electrostatic potential may have sharp gradients as long as they
produce a slowly varying $\eta$. To quantify what we mean by a sharp
gradient, we may balance the $v_\parallel \vec{b}\cdot \nabla g_a$ and
$\vec{v}_{ma} \cdot \nabla \psi \p_\psi f_{Ma}$ terms
in \eqref{eq:dke} to find that
\begin{equation}
\rho_{pa} |\nabla \psi| (\log X)' \ll 1
\label{eq:ordering} 
\end{equation}
should be satisfied by $T_a$ and $\eta_a$; that is, these quantities
should have a small relative change as experienced by a particle
during its radial drift excursion. Here $\rho_{pa}=v_a m_a /(e_a B_p)$
is the poloidal Larmor radius of the species, with the thermal speed
$v_a=\sqrt{2T_a/m_a}$, and the poloidal magnetic field
$B_p=\vec{B}\cdot \nabla \theta/|\nabla \theta|$.

\subsection{Model profiles and magnetic geometry}
\label{ssec:profiles}

Although we would like to use pedestal profiles which are
representative of experiments in some aspects, we
require \eqref{eq:ordering} to be satisfied in the simulations for all
species ($a=\{e,i,z\}$ for electrons, ions and impurities,
respectively). We consider pedestals where $n_a$ and $T_e$ are
allowed to vary on the $\rho_{pi}$ scale, while the $T$ and $\eta$
gradients of the bulk and impurity ion species are constrained
by \eqref{eq:ordering}. 

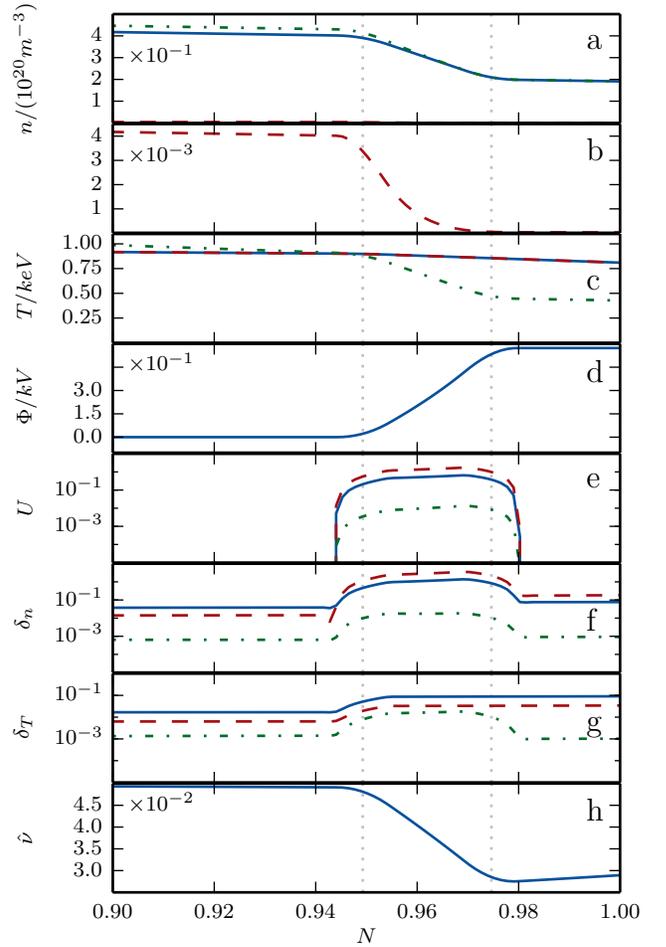
\begin{figure}
\includegraphics[width=1.0\columnwidth]{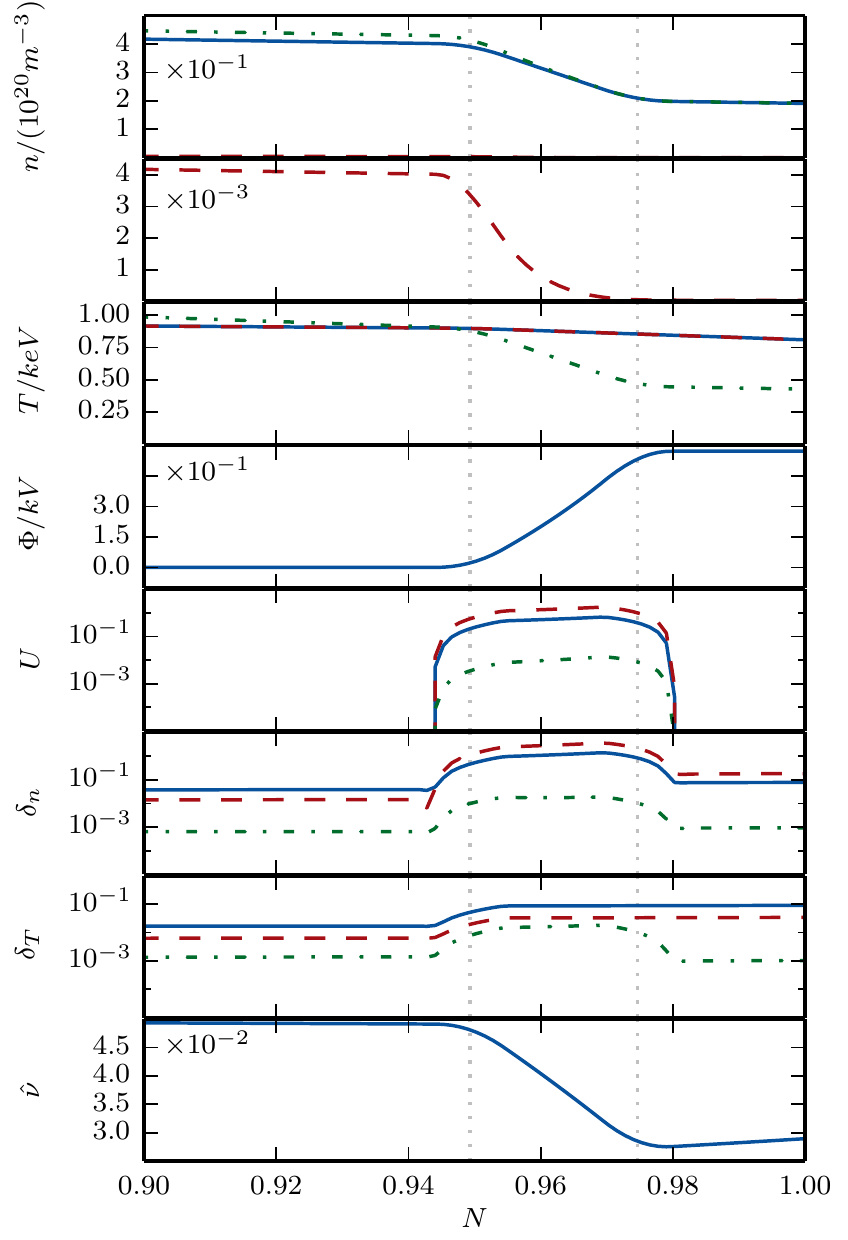}
\put(-25,342){\large a}
\put(-25,300){\large b}
\put(-25,251){\large c}
\put(-25,216){\large d}
\put(-25,176){\large e}
\put(-25,122){\large f}
\put(-25,84){\large g}
\put(-25,50){\large h}
\caption{\label{fig:baseline_in}  Ion (solid curve), impurity (dashed) 
and electron (dash-dotted) input profiles (a-d) and derived quantities
(e-h) for the baseline simulation.  }
\end{figure}

 As a starting point we considered profiles from typical JET
discharges (Figure 16 of \citenum{0029-5515-55-11-113031}), and
modified them as necessary to satisfy our orderings, and to reduce the
need for heat sources in the simulation domain. The specific choices
made in constructing the model profiles are discussed in
Appendix~\ref{sec:profilesapp}. The resulting input profiles for our
baseline case as functions of the normalized poloidal flux $\psi_N$
are shown in \autoref{fig:baseline_in}a-d. Here, we introduced
$\psi_N=\psi/\psi_{\rm LCFS}$, where $\psi_{\rm LCFS}$ is the poloidal
flux at the last closed flux surface (LCFS). Note that the density
profile of impurities is much steeper than that of the bulk ions to
make $\eta_z$ vary slowly. We consider a deuterium plasma and fully
ionized nitrogen impurities ($Z_z=7$, where $Z_a=e_a/|e_e|$) with a concentration
$n_z/n_i=0.01$ in the core. The radii marking the beginning and the
end of the pedestal are indicated by dotted vertical lines in the
figures.

In the simulations we use a local Miller model geometry
\cite{miller98} (and neglect the radial variation of $B_p$ and
$B_t=\sqrt{|\mathbf{B}|^2-B_p^2}$ in the domain), with elongation
$\kappa=1.58$, $s_\kappa\equiv (r/\kappa)\d\kappa/\d r=0.479$,
triangularity $\delta = 0.24$, $s_\delta\equiv
(r/\sqrt{1-\delta^2})\d\delta/\d r=0.845$, $\p R/\p r =-0.14$,
$q=3.5$, and inverse aspect ratio $\epsilon\equiv r/R=0.263$, with $r$
and $R$ denoting the minor and major radii, respectively.  These
parameters were taken from \citenum{belli2008}.

The numerical resolution and convergence tests are detailed
in Appendix~\ref{sec:resolapp}.

\subsection{Units}
 An input quantity $X$ is supplied to \perfect{} in a normalized,
dimensionless form $\hat{X}=X/\bar{X}$, with the normalizing,
dimensional quantity $\bar{X}$. We choose $\bar{R}=\unit[3.8]{m}$ and
$\bar{B}=\unit[2.9]{T}$ to be the major radius and magnetic field at
the magnetic axis. Furthermore we choose the following convenient
units $\bar{n}=\unit[10^{20}]{m^{-3}}$,
$\bar{T}=e\bar{\Phi}=\unit[1]{keV}$, $\bar{m}=m_D$ (deuterium mass),
and $\bar{e}$ is the elementary charge. We define the reference speed
and collision frequency as $\bar{v}=\sqrt{2\bar{T}/\bar{m}}$ and
\begin{equation}
\bar{\nu} = \frac{\sqrt{2}}{12 \pi^{3/2}}  
\frac{e^4\bar{n}\ln\Lambda}{\epsilon_0^2\sqrt{\bar{m}}\bar{T}^{3/2}},
\label{nubar}
\end{equation}
where $\epsilon_0$ denotes the vacuum permittivity and $\ln \Lambda$
is the Coulomb logarithm. The ordinary same-species collision
frequency $\nu_{aa}$ is defined as \eqref{nubar} but with $\bar{T}$
and $\bar{n}$ replaced by $T_a$ and $n_a$, from which we may define
the collisionality as $\hat{\nu}_{aa} = \nu_{aa} qR/v_a$.

To quantify when the local approximation fails, it is useful to define
a few additional quantities: The normalized electric field
$U_a=I/(Bv_a)\, \d\Phi_0/\d\psi$ measures how much the $E\times B$
drift competes with parallel streaming in terms of poloidal particle
motion. The normalized logarithmic derivative $\delta_X=-I
v_a/(\Omega_a)\, \d(\ln X)/\d\psi$ measures the variation of quantity
$X$ along a typical drift orbit, where $\Omega_a=e_a B/m_a$ and
$I=RB_t$.  Local neoclassical theory is valid only when $U$,
$\delta_n$, $\delta_\eta$, and $\delta_T$ are all much smaller than
unity in magnitude. The global $\delta f$ model also requires that
$|\delta_T|\ll 1$ and $|\delta_\eta|\ll 1$, but $U$ and $\delta_n$ can
be $\mathcal{O}(1)$.  These derived quantities together with
$\hat{\nu}$ are shown for the baseline profiles
in \autoref{fig:baseline_in}e-h.


\section{\label{sec:simulations} Local and global simulation results}

\def\colormap{red to violet through the colors of the rainbow}

To study the differences between local and global neoclassical
transport we performed a number of \perfect{} simulations with the
profiles and magnetic geometry described in \autoref{ssec:profiles}.
Before presenting the simulation results for our baseline set of
profiles, we discuss the output quantities.

We define the following normalized output quantities: sources,
$\hat{S}_a = \bar{v}^2\bar{R} S_a/(\Delta\bar{n}\hat{m}_a^{3/2})$, with $\Delta
= \bar{m}\bar{v}/(\bar{e}\bar{B}\bar{R})$;
$\hat{V}' = V' \bar{B}/\bar{R}$, with $V'=\oint d\theta
/|\vec{B}\cdot\nabla \theta|$; particle flux, $\hat{\vec{\Gamma}}_a =
\int d^3v g_a \vec{v}_{ma}/(\bar{n}\bar{v})$;
momentum flux (divided by mass), $\hat{\vec{\Pi}}_a =
\int d^3v g_a v_\parallel
I \vec{v}_{ma}/(\bar{n}\bar{v}^2\bar{R} B)$; heat flux,
$\hat{\vec{Q}}_a =\int d^3v g_a m_a
v^2\vec{v}_{ma}/(2 \bar{T}\bar{n}\bar{v})$; conductive heat fluxes,
$\hat{\vec{q}}_a =\hat{\vec{Q}}_a - (5/2)\hat{T}\hat{\vec{\Gamma}}_a$;
parallel flow velocity, $\hat{V}_{\parallel a} = \int d^3v v_\parallel
g_a/(\bar{v}\Delta n_a)$; parallel current, $\hat{J}_\parallel= \sum_a
Z_a \hat{n}_a\hat{V}_{\parallel a}$.  The neoclassical flow
coefficient $k_\|$ is defined so that it reduces to the flux function
poloidal flow coefficient for a single ion species plasma in the local
limit,
\begin{equation}
k_\parallel = \left(\frac{\d T}{\d\psi}\right)^{-1}\frac{\langle
B^2 \rangle}{B^2} \left(\frac{e_aB}{I} V_{\parallel,a}
+ \frac{1}{n_a} \frac{\d p_a}{\d\psi} +
e_a \frac{\d\Phi}{\d\psi}\right),\label{eq:kpar}
\end{equation}
where $p_a=n_aT_a$ is the pressure.  Furthermore, we define the
non-adiabatic density perturbation $\tilde{n}_a^{(g)}=\int d^3 v g_a$,
the total density perturbation $\tilde{n}_a=\int d^3 v (f_a-f_{Ma})$,
and normalized scalar fluxes
\begin{equation}
\begin{aligned}
\hat{F}_a = \frac{\hat{\psi}_a\hat{V}'\bar{R}}{\Delta^{2}\pi} 
  \langle \hat{\vec{F}}_a \cdot \nabla \psi_N\rangle, 
\end{aligned}
\label{scalarfluxes}
\end{equation}
with $\hat{F}_a$ ($\hat{\vec{F}}_a$) representing $\hat{\Gamma}_a$,
$\hat{q}_a$ or $\hat{\Pi}_a$ ($\hat{\vec{\Gamma}}_a$,
$\hat{\vec{q}}_a$ or $\hat{\vec{\Pi}}_a$, respectively), and
$\hat{\psi}_a=\psi_{\rm LCFS}/(\bar{R}^2\bar{B})$.  Note that the flux
normalizations are species independent.

\subsection{Results for the baseline case}

\begin{figure}
    \includegraphics{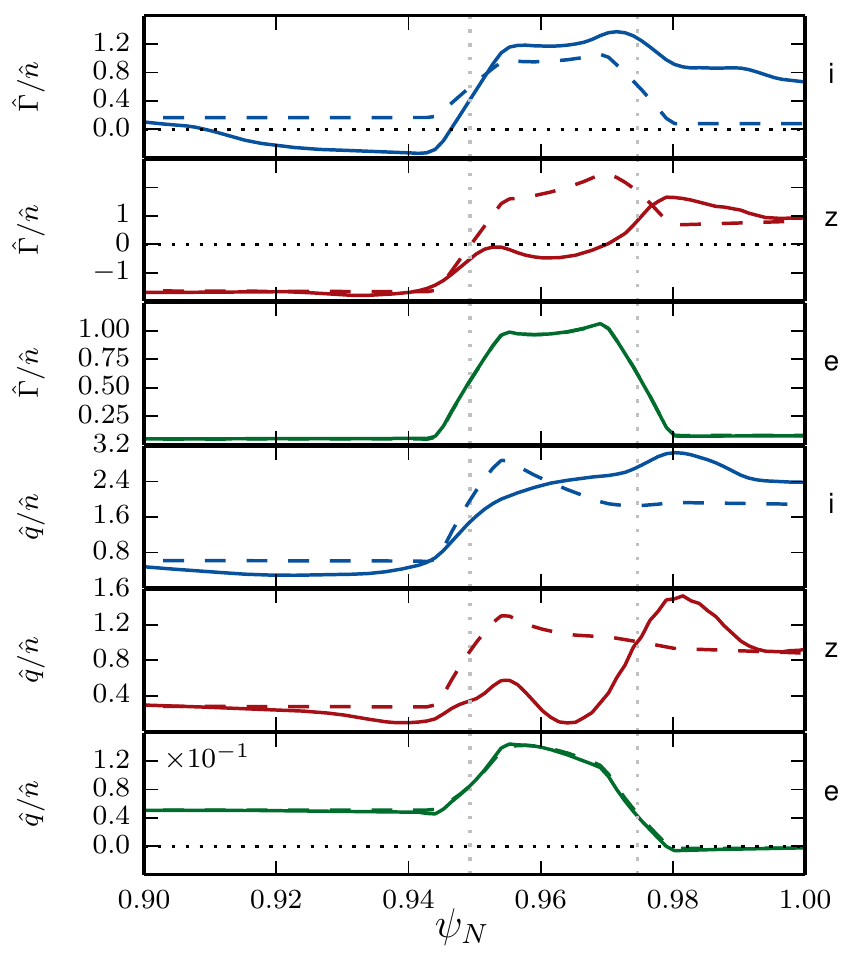}
\put(-25,258){\large a}
\put(-25,216){\large b}
\put(-25,176){\large c}
\put(-25,118){\large d}
\put(-25,92){\large e}
\put(-25,50){\large f}
\caption{\label{fig:fluxon} Particle (a-c)  and heat fluxes (d-f) divided by the
  normalized density, where the different subplots show the fluxes for
  different species. Solid (dashed) lines represent global (local)
  simulations. Species is indicated to the right of the panels.}
\end{figure}
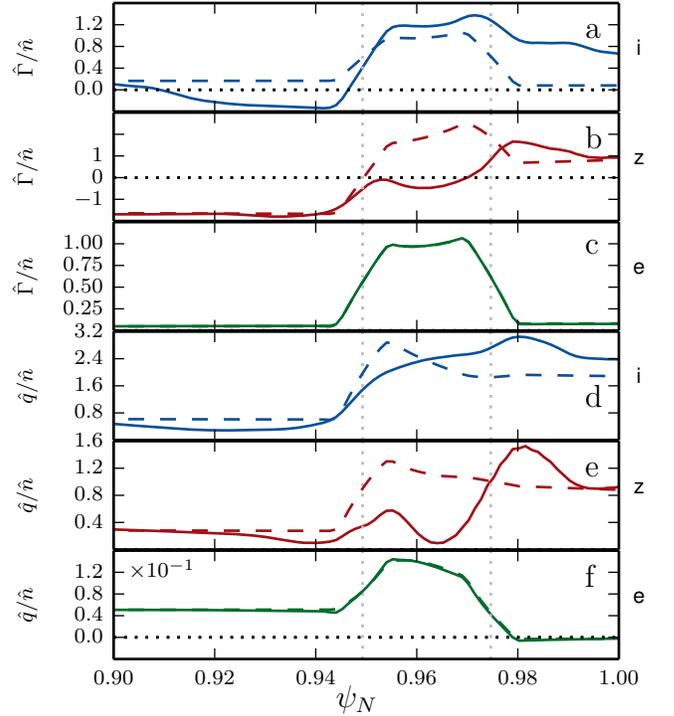

The scalar fluxes of \eqref{scalarfluxes} divided by $\hat{n}$ are
shown in \autoref{fig:fluxon}.  Throughout \autoref{sec:simulations}
solid lines represent global simulation results, and dashed lines
represent local ones.

When the impurity strength $\alpha=Z_{\rm eff}-1$ is order unity -- as
in our baseline -- the magnitude of the particle transport of
electrons is typically $\sqrt{m_e/m_i}/\alpha$ smaller than that of
the impurities, where $Z_{\rm eff}$ denotes the effective ion
charge. In this case it is common to neglect $\hat{\Gamma}_e$ and
calculate the ion particle transport from ambipolarity $\sum_{a \ne e}
Z_a\hat{\Gamma}_a=0$, leading to opposing ion and impurity particle
fluxes. The local simulations in the core region obey these
expectations. Since $\d T_i/\d\psi$ is small in our baseline,
temperature screening does not dominate, thus the local fluxes obey
$\hat{\Gamma}_i>0$ and $\hat{\Gamma}_z<0$. In the pedestal the
parallel ion-electron friction can be sufficiently large to compete
with the ion-impurity friction, due to the high electron flow
speeds. Therefore $\hat{\Gamma}_e$ cannot be neglected anymore in the
ambipolarity condition: the strong outward electron flux means that both
$\hat{\Gamma}_z$ and $\hat{\Gamma}_i$ are positive simultaneously.
Thus the outward local $\hat{\Gamma}_z$ is not a result of temperature
screening.

As expected from the small $U_e$ and $\delta_{ne}$ values, seen in
\autoref{fig:baseline_in}e and f, the electron local and global fluxes
are practically the same. However, finite orbit width effects strongly
affect the ion and impurity dynamics. In the pedestal $\hat{\Gamma}_i$
is increased compared to the local value, which causes
$\hat{\Gamma}_z$ to change sign compared to its local value. It is
worth noting that the deviation between local and global results is
not localized to the pedestal region only. For instance, the global
and local $\hat{\Gamma}_i$ deviate well below $\psi_N=0.94$; the ion
particle flux changes sign at $\psi_N=0.91$ while the local result is
positive everywhere.  As a comparison we note, that the width of
  the large gradient region in $\psi_N$ units is approximately
  $0.026$, and the orbit width of a typical trapped ion at thermal
  speed is $0.016$. 

The somewhat surprising observation that the finite orbit width
effects extend outside the pedestal over several thermal ion orbit
widths is worth a moment of thought. Since the existing analytical
theories assume $\sqrt{\epsilon}\ll 1$, which partly eliminates the
radial coupling, they can only provide a limited guidance as to why
this happens. The only radial coupling that cannot be completely
eliminated from those theories is that due to the neoclassical
parallel flow, $k_\|$; indeed it is not a parameter in the theory, but
it satisfies a radial differential equation (Eqs.~(43) and (64) in
Ref.~\citenum{catto13}). As we will see, in our case the global result
for the main ion $k_\|$ is very different from the local one, and it
takes a rather long distance from the pedestal before it gets close to
the local result. Also, in estimating the orbit width above, we
considered particles at the thermal speed, while all the quantities of
interest are dominated by super-thermal particles with wider
orbits. The neoclassical drive and the radial coupling terms both
include $\vec{v}_m \cdot \nabla \psi\propto v^2$, the velocity space
integration weight is $\propto v^2$, and the flow, particle flux, and
heat flux contain an additional factor of $v$, $v^2$, or $v^4$
respectively.

 The global particle fluxes are not ambipolar as seen at
 $\psi_N=0.94$, where $\hat{\Gamma}_i$ and $\hat{\Gamma}_z$ are both
 inward, and $\hat{\Gamma}_e$ is small, very close to its local value.
 Note that while in a local simulation the radial current and momentum
 flux should vanish, it does not need to be so in global simulations,
 as pointed out in \citenum{landreman2014}.

For both ions and impurities we observe that the conductive heat flux
can significantly differ from the local value. A reduction compared to
the the local value -- observed around the pedestal top -- may be
explained by the shift of the trapped region towards the tail of the
distribution at $U\sim 1$.  However, we also find regions where the
heat fluxes increase from their local values. Just as for the particle
fluxes, we see a reduction inside the pedestal top: $\hat{q}_i$
($\hat{q}_z$) reaches a minimum at $\psi_N\approx 0.92$
($\psi_N\approx 0.94$).

\begin{figure}
    \includegraphics[width=1.0\columnwidth]{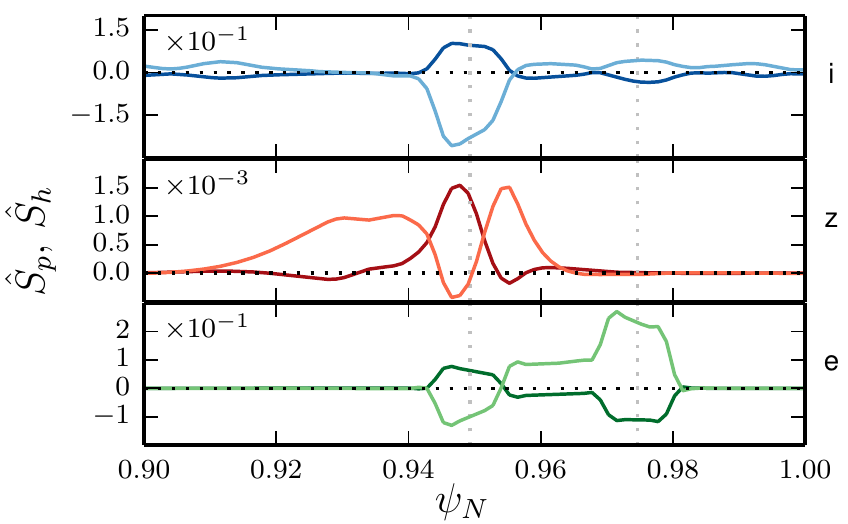}
\put(-25,118){\large a}
\put(-25,92){\large b}
\put(-25,50){\large c}
\caption{\label{fig:sources} Particle (dark curves) and heat (light
  curves) sources for the various species. Species is indicated to the right of the panels.}
\end{figure}

 The corresponding sources are presented in \autoref{fig:sources}. We
use poloidally symmetric sources with speed dependencies
$(x_a^2-\{5,3\}/2)\exp(-x_a^2)$ for particle and heat sources,
respectively, where $x_a=v/v_a$. We see that to some degree the main
ion particle and heat sources qualitatively mirror each other, and
tend towards zero outside the pedestal. We specifically
choose $T_i$ to reduce the need for $S_{hi}$, while $S_{pi}$ also
remains small, because $\hat{\Gamma}_i$ tends to be smaller than
$\hat{q}_i$. The tremendous drop in $\hat{n}_z$ in the pedestal leads
to a sharp peak in $S_{pz}$, and again we see an opposing trend for
$S_{hz}$, but for impurities the combined sources are positive. The
electron sources are localized to the pedestal and comparable in size
to those of the main ions.

\begin{figure}
    \includegraphics[width=1.0\columnwidth]{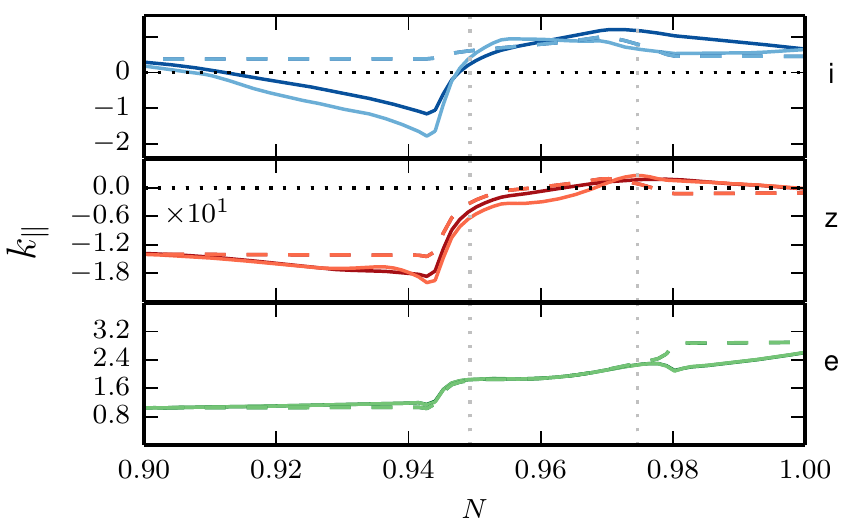}
\put(-25,118){\large a}
\put(-25,80){\large b}
\put(-25,38){\large c}
\caption{\label{fig:kPar} $k_\parallel$, defined in \eqref{eq:kpar}, with
 darker lines being $k_\parallel$ at the inboard side, $\theta=\pi$, lighter 
at $\theta=0$; dashed lines show the results of corresponding local 
simulations (the outboard and inboard curves overlap). Species is indicated to the right of the panels.}     
\end{figure}

The neoclassical flow coefficients, $k_\parallel$, are presented in
\autoref{fig:kPar}. The local (dashed lines) $k_{\| i}$ is positive as
expected in the banana regime, and exhibits a slight variation as a
response to the radial variation of $\hat{\nu}$. In the local case
$k_\|$ is a flux function, while globally it varies from the inboard
side (darker curves) to the outboard side (lighter curves). These
poloidal variations in the flow appear together with poloidal density
variations, as will be discussed shortly. It has been shown
analytically \cite{kagan10PPCF, pusztai10, catto13} that $k_\|$ is
affected by finite orbit width effects. As seen from the analytical
results, where $\vec{v}_{da0}\cdot \nabla\theta$ is kept but the
radial coupling is neglected, $k_\|$ is expected to decrease or become
more negative in both the banana and plateau regimes
\cite{kagan10PPCF,pusztai10}. However, the global modification to
$k_\|$ is not a function of local plasma parameters (such as $U_a$)
only, but even in the semi-global treatment of \citenum{catto13} it
satisfies a radial differential equation (i.e. $\partial
g/\partial\psi$ cannot be neglected). This is why $k_\|$ can differ in
sign and magnitude from the local value well outside the pedestal, and
can be larger than its local value inside the pedestal. It is
interesting to note that there is a difference between global and
local $k_\|$ even for the electrons, which is due to the collisional
coupling to the various ion species.

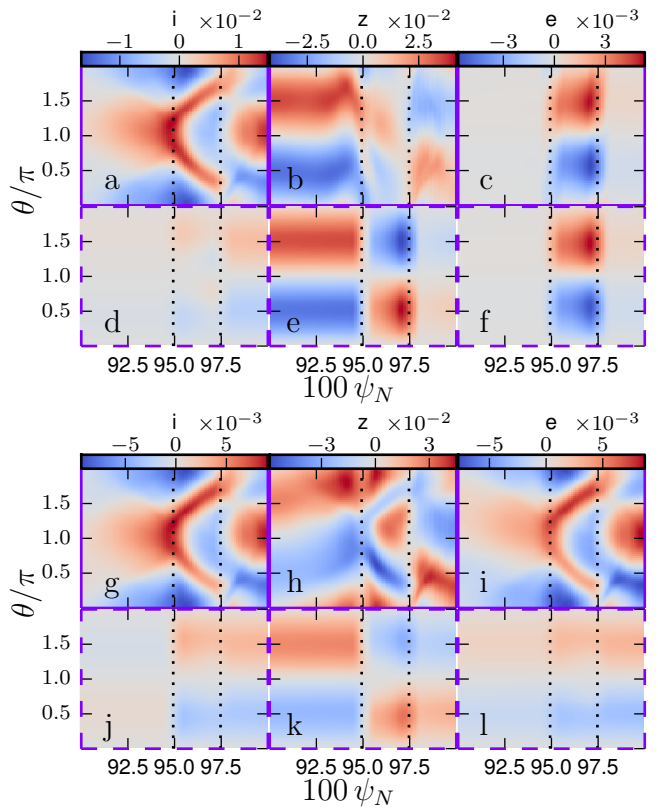
\begin{figure}
\begin{subfigure}[b]{1.0\columnwidth}
    \includegraphics{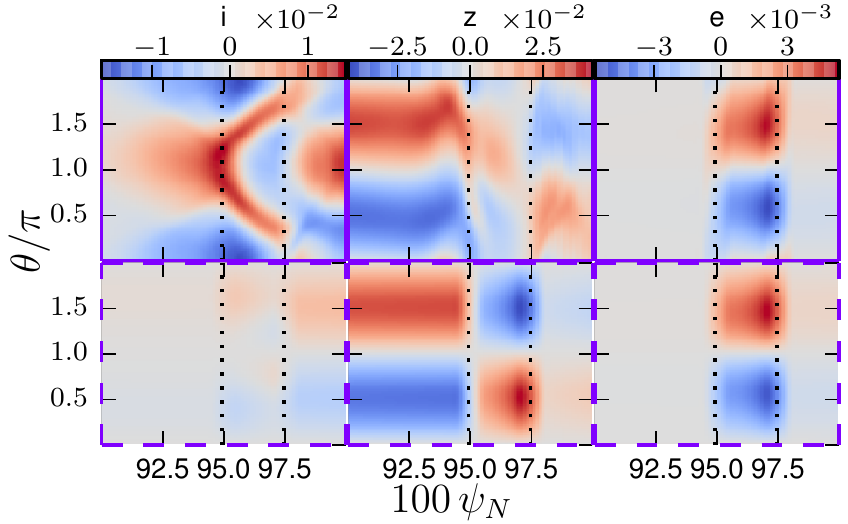}
\put(-207,82){\large a}
\put(-138,82){\large b}
\put(-65,82){\large c}
\put(-207,28){\large d}
\put(-138,28){\large e}
\put(-65,28){\large f}   
  \end{subfigure}
\begin{subfigure}[b]{1.0\columnwidth}
    \includegraphics{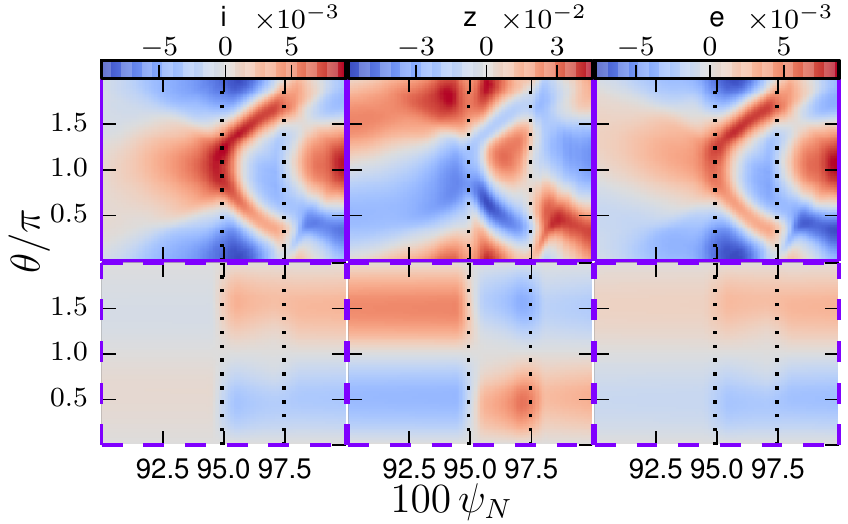}
\put(-207,82){\large g}
\put(-138,82){\large h}
\put(-65,82){\large i}
\put(-207,28){\large j}
\put(-138,28){\large k}
\put(-65,28){\large l}
\end{subfigure} 
\caption{\label{fig:baselinDensPert} The non-adiabatic contribution to the 
density perturbation (a-f), and the total
  density perturbation (g-l). Dashed frame (d-f and j-l) indicates
  local results. Species is indicated above the panels.}
\end{figure}

The density perturbations are shown in
\autoref{fig:baselinDensPert}. We first consider only the
non-adiabatic contribution to the perturbations,
$\tilde{n}_a^{(g)}/n_a$, shown in \autoref{fig:baselinDensPert}a-f,
where the global (local) results are plotted in the a-c (d-f)
panels. The local simulations predict purely up-down density asymmetry
for all species, which is weaker for bulk ions and electrons than for
impurities.  The global results show a more complex poloidal density
variation for both ions and impurities. For ions we see an in-out
asymmetry (i.e. excess density around $\theta=\pi$) at the pedestal
top, which transforms into an out-in asymmetry in the pedestal, and
reverses again further out (similarly to the single species
simulations of \citenum{pusztaiRFNEO}). For impurities the most
important difference compared to the local results is the weak in-out
(instead of strong up-down) asymmetry in the pedestal. The electron
density perturbation mostly follows its local behavior, exhibiting a
large increase in the up-down asymmetry in the pedestal. To understand
the total density perturbation $\tilde{n}_a/n_a$ in global simulations
we note that the potential perturbation $\Phi_1$ follows mostly the
non-adiabatic ion density perturbation. The total electron density
perturbation is dominated by the adiabatic response of electrons, thus
it is very similar to the ion density perturbation.  The impurity
density variations show a competition between adiabatic response --
especially in the pedestal where it tries to oppose the ion density
perturbation -- and non-adiabatic response. The relative impurity
density variation stays below $10\%$ everywhere, showing that the
assumption of the density being nearly a flux function is not
violated. However, for sufficiently high $Z_z$, nonlinearity from
poloidal asymmetries can arise \cite{fulop99,fulop01,landreman11}.

\begin{figure}
  \includegraphics[width=1.0\columnwidth]{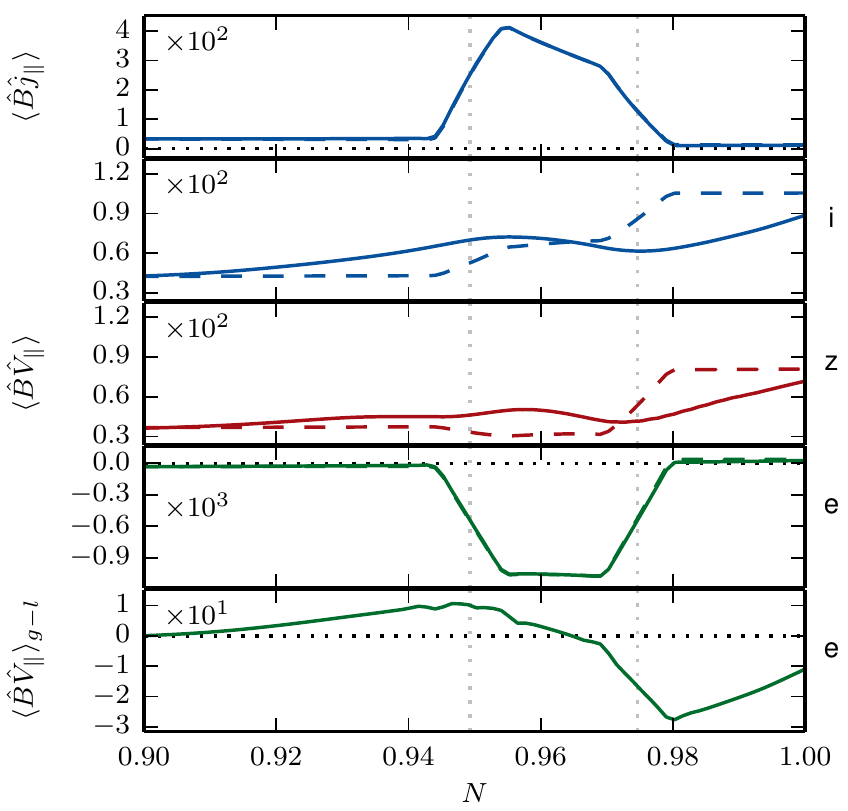}
 \put(-25,210){\large a} 
\put(-25,155){\large b}
\put(-25,112){\large c}
\put(-25,74){\large d}
\put(-25,40){\large e}
\caption{\label{fig:FSAVPar} $\langle BV_\parallel \rangle$ for the
  different species (b-d), and the parallel current these add up to
  (a).  Solid (dashed) lines represent global (local)
  simulation. The difference between the global and local
    electron parallel flows is shown in (e). Species is indicated to the right of the panels.}
\end{figure}
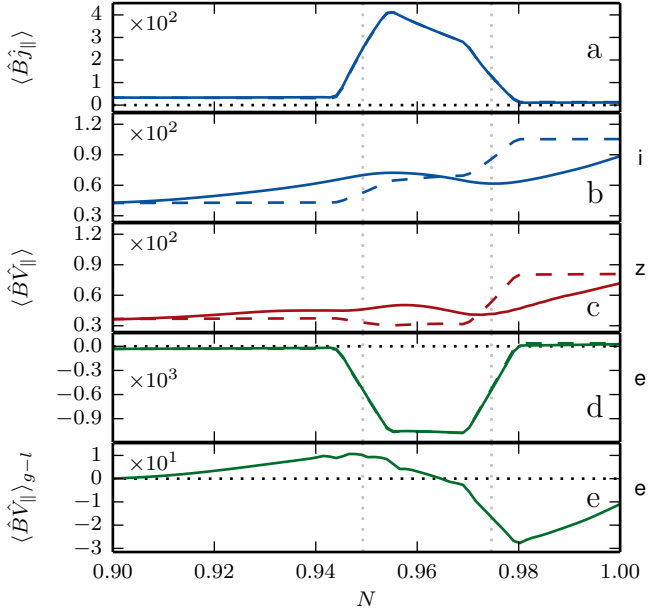

Finally, we consider the parallel flows and the bootstrap current for
our baseline case; these are plotted in \autoref{fig:FSAVPar}.  For
the bulk and impurity ions the flows remain small inside the pedestal,
as their profiles were chosen specifically so that their diamagnetic
and $E\times B$ flows mostly cancel. For these the relative deviation
between the local and global results is significant, the global
results being larger in magnitude from the middle of the pedestal
inward. This is partly due to the reduction in $k_\|$ compared to the
local value observed in \autoref{fig:kPar}. The parallel flow of
electrons reaches a much higher magnitude inside the pedestal then
that of the ions, where the strong $E\times B$ and diamagnetic
rotation contribute with the same sign for this species. The scale is
therefore different and the difference between the local and global
results is less visible. Figure~\autoref{fig:FSAVPar}e shows this
difference, which is comparable to what is observed for ions. This is
expected, because all the difference is due to a frictional coupling
to the various ion species with modified flow speeds, as direct finite
orbit width effects are negligible for electrons. Since we have a weak
ion temperature pedestal, the modifications of the ion flows are not
sufficient to cause an appreciable deviation of the bootstrap current
from the local result inside the pedestal, as seen in
\autoref{fig:FSAVPar}a. If anything, outside the pedestal there is a
slight difference between the local and global results (barely visible
on the scale accommodating the huge bootstrap peak in the pedestal),
since the electron flow is relatively small in those regions, while
the global effects on $k_\|$ extend outside the pedestal.

\subsection{Non-trace impurities}
We want to assess whether, and how, non-trace impurities can affect
collisional transport in the pedestal. To this end, we performed
simulations with profiles similar to our baseline, except that we
scaled the impurity concentration profiles. The simulations
shown in this section have impurity concentrations such that the
impurity strength in the core is either $\alpha=0.0171$ (trace
impurities; shown with thin lines in the figures) or
$\alpha=1.33$ (non-trace; thick lines); as a comparison
  $\alpha=0.3925$ for the baseline. Note that $n_z/n_i$ drops rapidly
across the pedestal, thus the impurities eventually become trace even
if they have a high core concentration (this is necessary in the
presence of an electron density pedestal if we restrict profile
variations to obey the orderings of \eqref{eq:ordering} for all ion
species).

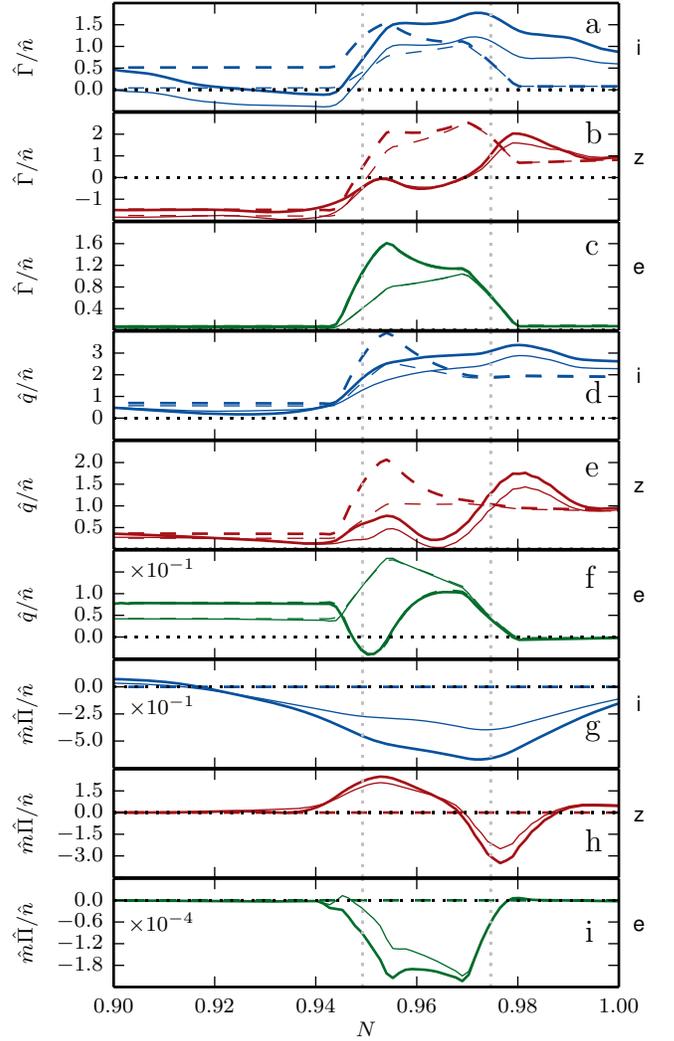
\begin{figure}
    \includegraphics{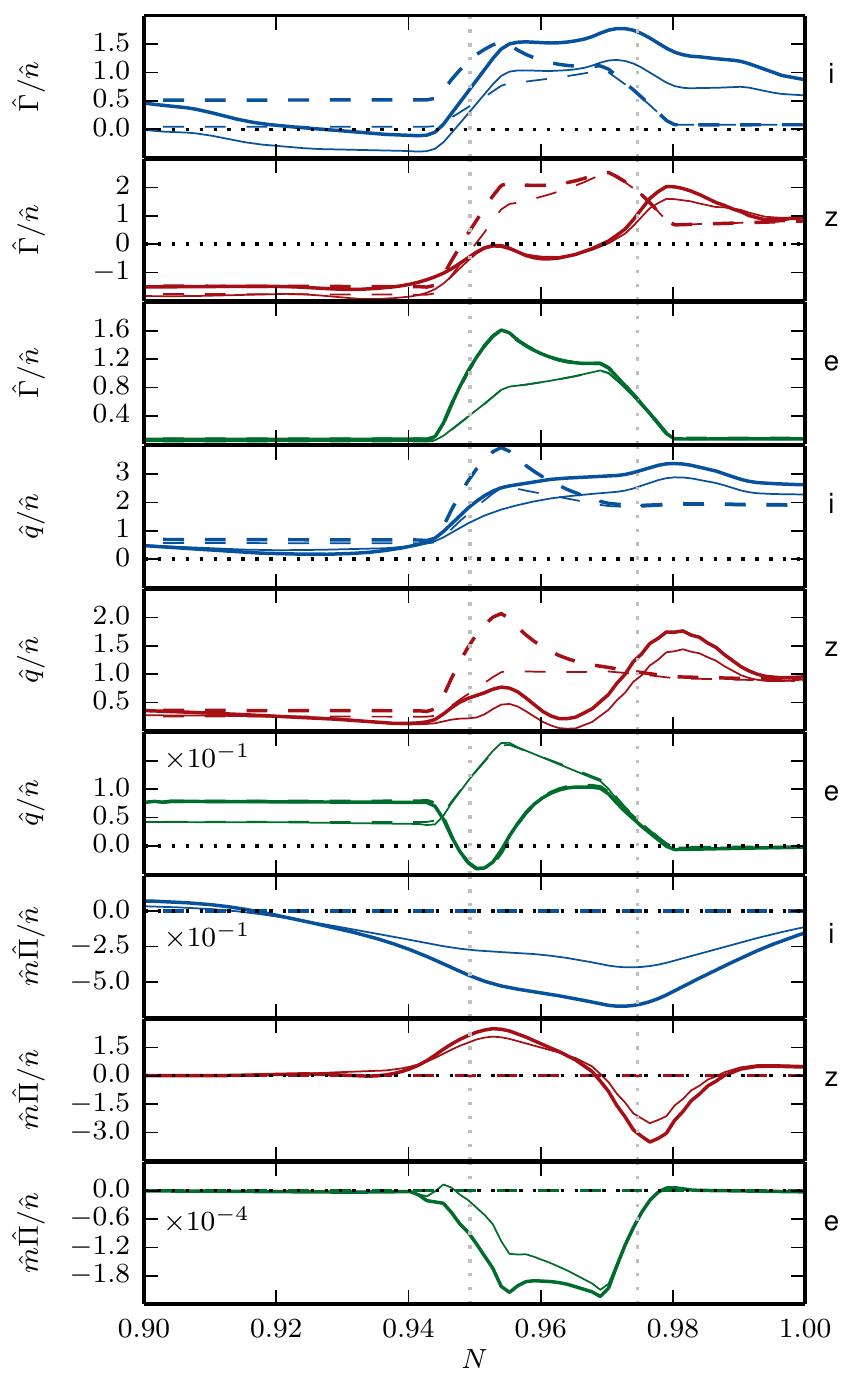}
\put(-25,384){\large a}
\put(-25,342){\large b}
\put(-25,300){\large c}
\put(-25,244){\large d}
\put(-25,216){\large e}
\put(-25,174){\large f}
\put(-25,118){\large g}
\put(-25,74){\large h}
\put(-25,40){\large i}
\caption{\label{fig:conc_fluxon} Particle (a-c), heat (d-f) and
  momentum (g-i) fluxes divided by the normalized density for the
  various species. Solid (dashed) lines represent global (local)
  simulations, and thick (thin) lines correspond to $\alpha=1.33$
    ($\alpha=0.0171$). Species is indicated to the right
    of the panels.}
\end{figure}

The a-c panels of \autoref{fig:conc_fluxon} compare the local and
global particle fluxes with trace and non-trace impurity
concentration. As usual, the local and global results for
$\hat{\Gamma}_e$ are virtually the same. At higher impurity content
the increase in $\hat{\Gamma}_e/\hat{n}_e$ around $\psi_N=0.95$
reflects the increased $n_e$ gradient due to the rapidly varying
impurity concentration (note that $n_i$ is kept fixed in the impurity
scan, thus the $n_e$ profile changes). The local $\hat{\Gamma}_i$
behaves as expected from the ambipolarity condition: it increases with
impurity content in the core to balance the inward $\hat{\Gamma}_z$,
and its core behavior is dominated by following the outward electron
flux. It is interesting to note that while above $\psi_N=0.97$ the
local curves for different impurity concentration collapse onto each
other because of the low $n_z/n_i$, the difference in the global
$\hat{\Gamma}_i/\hat{n}_i$ survives much further out in the pedestal.

The d-f panels of \autoref{fig:conc_fluxon} show the heat fluxes.  In
the core and close to the pedestal top the global results for $q_i$
are lower than the local values. This reduction is somewhat stronger
in the presence of impurities. The local results for $\hat{\Gamma}_z$
and $\hat{q}_z$ are even higher in the pedestal at high impurity
concentration, than would be expected simply due to the linear
increase with $\hat{n}_z$. This extra increase is reduced by the
global effects so that the global results at different concentrations
are closer to each other.

\begin{figure}
    \includegraphics{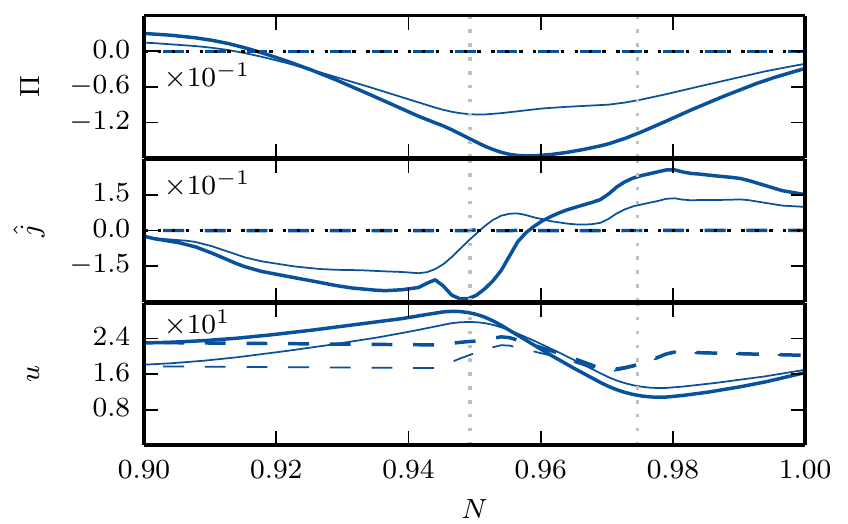}
\put(-25,116){\large a}
\put(-25,72){\large b}
\put(-25,30){\large c}
\caption{\label{fig:conc_momsum}
Total momentum flux (a), radial current (b) and parallel mass flow
(c). Thick (thin) lines correspond to $\alpha=1.33$ ($\alpha=0.0171$).}
\end{figure}

The momentum flux values for the different species (shown
in \autoref{fig:conc_fluxon}g-i) vanish in the local limit, as they
should, but are finite when global effects are considered. Unlike
$\hat{\Gamma}_e$ and $\hat{q}_e$, the global results for $\hat{\Pi}_e$
are different to the local ones, i.e. they are also finite, although
the electrons do not transport an appreciable amount of momentum, as
$m_e/m_i\ll 1$. The total momentum flux,
$\Pi=\sum_a \hat{m}_a\hat{\Pi}_a$, shown
in \autoref{fig:conc_momsum}a, is mostly negative in the studied
radial range, and is strongly increase by the presence of non-trace
impurities.  The ions are responsible for most of the momentum
transport for both impurity concentrations, thus the increase in
$|\Pi|$ is not due to the increase in $|\hat{\Pi}_z|$, but the
modifications in $\hat{\Pi}_i$ in the presence of impurities.

 The radial current $\hat{j}=\sum_a Z_a \hat{\Gamma}_a$
(\autoref{fig:conc_momsum}b) -- that in isolation from other
transport channels would lead to charge separation and the evolution
of the radial electric field -- is also significantly increased in
magnitude by the presence of the impurities over most of the studied
radial domain.  Although we observe a finite radial neoclassical
current, we do not attempt to self-consistently calculate the radial
electric field. For the interpretation of the momentum fluxes and the
radial current it is useful to note that conservation of particle
number and parallel momentum (in steady state, for a radially constant
$V'$, and with sources even in $v_\|$) imply that
$\d\hat{\Gamma}_a/\d\psi_N\propto \hat{T}^{3/2}\hat{S}_{pa}$ and
$\d\Pi/\d\psi_N\propto \hat{j}$. The latter property is apparent from
a comparison of \autoref{fig:conc_momsum}a and b. The former relation
states that non-ambipolar fluxes require particle sources for which
$\sum_a Z_a \hat{S}_{pa}= 0$ is not satisfied locally (the factor
$\hat{T}^{3/2}$ comes from the assumed velocity space structure of
$S_{pa}$).

\begin{figure}
    \includegraphics{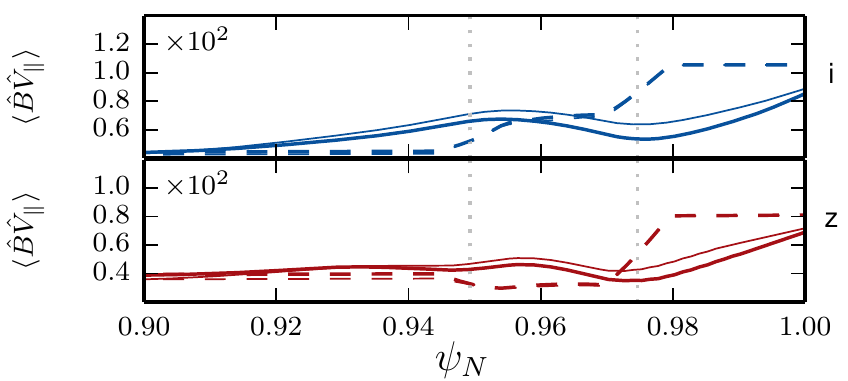}
\put(-25,96){\large a}
\put(-25,50){\large b}
\caption{\label{fig:conc_FSAVPar} $\langle BV_\parallel \rangle$ for
  ions and impurities. Solid (dashed) lines represent global (local)
  simulations, and thick (thin) lines correspond to $\alpha=1.33$
    ($\alpha=0.0171$). Species is indicated to the right
    of the panels.}
\end{figure}

In radially local formalisms momentum transport is often decomposed
into diffusive ($\propto - \d u_a/\d\psi_N$, where
$u_a=\hat{n}_a \langle \hat{B} \hat{V}_{\| a}\rangle \hat{m}_a$ is the
normalized parallel mass flow), conductive ($\propto u_a$) and
intrinsic (independent of $u_a$) terms. Such a decomposition is not
possible in our global formalism since the parallel mass flow is a
non-local function of the various plasma parameter profiles, and so is
the momentum transport. Nevertheless, it is instructive to compare the
radial profile of the total momentum transport $\Pi$ to the radial
variation of the total mass flow $u=\sum_a u_a$ shown
in \autoref{fig:conc_momsum}c (both quantities are dominated by the
main ion contributions). By increasing the impurity content, $u$
increases in the core region.  This is caused by the higher
$\hat{n}_z\langle\hat{B}\hat{V}_{\|z}\rangle$ at higher $\hat{n}_z$
not being compensated by the slight reduction in
$\langle\hat{B}\hat{V}_{\|i}\rangle$ (shown in
\autoref{fig:conc_FSAVPar}). As the impurity concentration drops radially
across the pedestal, the effect from the reduction in the parallel ion
flow becomes dominant. The radial drop in the global results for $u$
across the pedestal is mostly due to the density variation in the
pedestal. If the transport was local and purely diffusive this
non-monotonic behavior of $u$ would be accompanied by a sign change in
the momentum transport. It is also interesting to note that at the
point where we see the greatest relative increase in $|\Pi|$ between
the different simulations ($\psi_N\approx 0.96$), the global $u$
becomes lower for higher core impurity concentration.

\begin{figure}
    \includegraphics{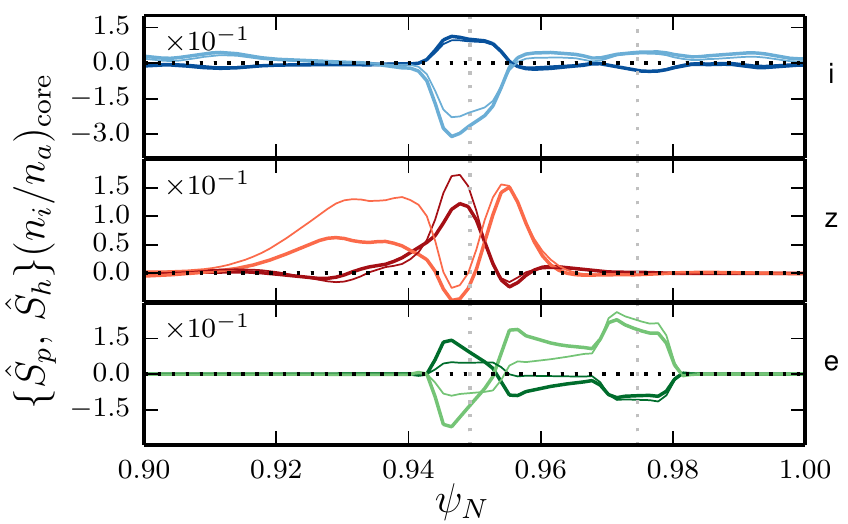}
\put(-25,118){\large a}
\put(-25,84){\large b}
\put(-25,50){\large c}
\caption{\label{fig:conc_sources} Particle (dark curves) and heat
  (light curves) sources normalized by the species concentration in
  the core. Thick (thin) lines correspond to $\alpha=1.33$
  ($\alpha=0.0171$). Species is indicated to the right of the panels.}
\end{figure}

The ion particle source profile shown in
\autoref{fig:conc_sources}a is only weakly affected by the presence of
non-trace impurities. In the meantime, the increase in impurity
sources are approximately proportional to the increase in their
concentration: the normalized particle source
$\hat{S}_p(n_i/n_z)|_{\rm core}$ is approximately the same in the two
simulations, considering that the impurity content changes by a factor
$100$.  

The radial current and the non-quasineutral particle sources $S_{pa}$
are consistent, that is, the divergence of the radial current is given
by the charge source $\sum_a Z_aS_{pa}$ (note, that the total charge
source integrated across the pedestal is zero, due to the boundary
conditions). However, the radial current is truly a consequence of the
radial coupling in the global simulation, and not an artifact of the
radially varying sources. Although it is not done in the code, source
profiles could be calculated in the presence of the radially varying
\emph{local} particle and heat fluxes, for these to be consistent with
the time independent plasma parameter profiles. Such sources would be
quasineutral unlike those in the global simulations.

In reality, the neoclassical radial current that we observe needs to
be balanced by an opposing radial current, which represents transport
processes not captured by our model (turbulence, atomic physics
processes, orbit losses, magnetic ripple effects, etc.). Otherwise the
system would not be steady state, because the radial electric field
would vary in time and the $\mathbf{j}\times\mathbf{B}$ torque would
change the plasma flows. Due to the construction of the code and the
vanishing $v_\|$ moment of our sources, we observe the $(1/V')d (V'
\Pi)/d\psi $ and the radial current terms in the species-summed flux
surface averaged angular momentum equation to exactly
balance. Similarly, the finite neoclassical momentum transport
predicted by the code should also be canceled by a momentum transport
due to non-neoclassical processes, in steady state.

\begin{figure}
    \includegraphics{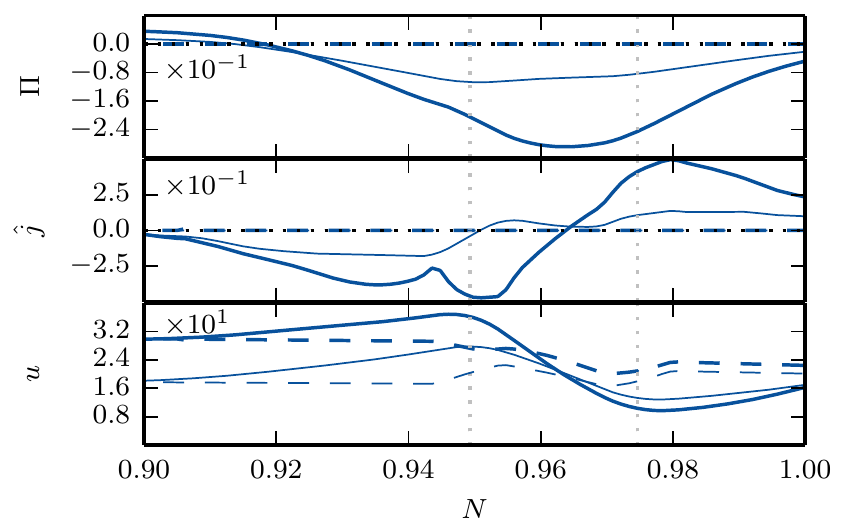}
\put(-25,120){\large a}
\put(-25,78){\large b}
\put(-25,50){\large c}
\caption{\label{fig:Z4conc_momsum}
Total momentum flux (a), radial current (b) and parallel mass flow (c)
for beryllium impurity with modified $\eta$ profiles that make the
impurity density profile less steep. Thick (thin) lines correspond to
 $\alpha=1.33$ ($\alpha=0.0171$).}
\end{figure}

Finally we would like to assess how much our profile choices affect
the observation of increased momentum flux in the presence of
non-trace impurities. One unnatural feature of our impurity density
profile is its extreme steepness. Allowing $\eta_i$ have a radial drop
across the pedestal -- within what is allowed by \eqref{eq:ordering}
-- to reduce the inward radial electric field, together with allowing
$\eta_z$ to have an increase in the same region, leads to a less sharp
impurity pedestal. To further reduce the impurity density gradient in
the pedestal we consider a fully ionized beryllium impurity
($Z=4$). These changes make our pedestal less deeply sub-sonic, as
seen from the increased mass flow in \autoref{fig:Z4conc_momsum}c
that shows the results of the modified input profiles.
From \autoref{fig:Z4conc_momsum}a we can conclude that impurities also
significantly increase the magnitude of the neoclassical momentum flux
in the pedestal for more natural impurity profiles, when the inputs
push the limitations imposed by the required orderings.


\section{\label{sec:discussion} Discussion and conclusions}

We have studied the differences in the collisional transport between
radially local and global formalisms using the global $\delta f$
neoclassical solver \perfect, with a special emphasis on the effects
of non-trace impurities. If the impurity density profile is arbitrary,
impurities are likely to develop sonic flows and strong poloidal
asymmetries in the pedestal. We use model profiles specifically chosen
to guarantee that impurity flows remain sub-sonic, so that the
assumptions of the $\delta f$ formalism are valid. In this way we can
gain some insights into the effect of non-trace impurities without the
need for a non-linear collision operator.

In an impure plasma the electron particle transport is usually
negligible, and the ion and impurity fluxes oppose each other to
maintain ambipolarity. However, since in the pedestal the magnitude of
the parallel electron flow can be much larger than the parallel ion
and impurity flows, the friction of the various ion species on
electrons can become non-negligible. Consequently, a substantial neoclassical
electron particle transport can arise, which competes with the radial
transport of other species. In particular, the ion and impurity fluxes
can have the same sign. In the presence of strong radial profile
variations on the $\rho_{pi}$ scale, the ambipolarity of fluxes is
violated, as reported in previous studies \cite{landreman2014}. We
emphasize that the differences in neoclassical flows and fluxes
between the local and global theory are not restricted to the pedestal
region only but die off within a distance comparable with the ion
orbit width. In our baseline from the outer core region to the
pedestal top we observe that both the ion and impurity fluxes are
inward due to global effects (see \autoref{fig:fluxon}a-c). Inside the
pedestal, local simulations predict both the impurity and the ion
fluxes to be outward, due to the large outward electron flux, but in
global simulations the impurity flux reverses to be inward.

For our model profiles, chosen to give small parallel ion
flows, the bootstrap current remains almost completely unaffected,
while this need not be so if the ion temperature varies more rapidly
than allowed by our orderings. Indeed, the neoclassical parallel flow
coefficient $k_\|$ of ions and impurities is significantly affected by
global effects. Due to the radial coupling even the sign of $k_{\| i}$
is different from the local result well inside of the pedestal top.

We observe that in the presence of global effects the poloidal
variation of the density perturbation is not restricted to an up-down
asymmetry, it can be more significant, and can develop rapid radial
variations (in accordance with previous numerical
results \cite{landreman12,pusztaiRFNEO}).  The poloidal asymmetries
observed here arise in the presence of finite inverse aspect ratio
and finite orbit width effects, and are not closely related to those
predicted by the analytical theories in
\cite{fulop99,fulop01,landreman11}, since those require significant
poloidal variations in the ion-impurity friction.  The adiabatic
response to the electrostatic perturbation generated by the ions
competes with non-adiabatic contributions in determining the poloidal
asymmetries developed by the impurities. For the moderate impurity
charge considered here ($Z_z=7$) the relative poloidal variation of the
impurity density is still small so that the perturbative treatment
remains valid. At sufficiently high $Z$ the relative poloidal
variation of the impurity density is expected to become order unity
and then -- if the impurities are non-trace -- nonlinear effects would
start to play a role. The numerical investigation of that situation is
left for a future study.

In the pedestal, the neoclassical radial current does not vanish in
general, and the momentum transport remains finite -- in contrast to the
local theory.  In the vicinity of the pedestal the total neoclassical
momentum transport is found to be negative in the studied case, which
happens if the charge sources (resulting from the particle sources
needed to sustain the pedestal) are mostly positive in that
region. The radial variation of the parallel mass flow is
non-monotonic with a sharply decreasing feature where the density
drops in the pedestal. This non-monotonicity is not reflected in the
radial momentum transport, which is now a non-local function of the
mass flow. A simple decomposition of the momentum transport into
diffusive, conductive and intrinsic terms is not possible in the global
picture. Note that the same is also true for all the radial fluxes. 

We observe a strong effect of impurities on the total momentum
transport, the magnitude of which increases significantly in the
presence of non-trace impurities (see
\autoref{fig:conc_momsum}a). While it is not possible to disentangle
the exact cause of this, it may be due to the sharp radial variation
of the parallel impurity mass flow. The impurities represent only a
minor fraction of the total mass flow, but their collisional coupling
to the main ions is significant at an impurity strength of order
unity. Pushing the limitations of our orderings we reduced the
sharpness of the impurity density profile, in an attempt to
demonstrate the robustness of the impact of impurities on neoclassical
momentum transport. Although the impurity profile is chosen to be very
specific, strong effects may occur when the impurities have more
general density variations, especially because they then have much
stronger relative flow speeds compared to the ions.

Impurity seeding in tokamaks operating with ITER-like metallic walls
has been experimentally found to have beneficial effects on the
pedestal performance. It is then natural to raise the question of
whether our results are consistent with this observation. Without
taking into account other non-intrinsically ambipolar processes and
turbulent transport it is not possible to evolve the profiles towards
a steady state. However, we may speculate about possible consequences
of the increased momentum transport in the presence of impurities. In
steady state the neoclassical radial current and momentum transport
should be balanced by opposing contributions of turbulent and other
origin. An increased neoclassical momentum transport in the presence
of impurities requires these contributions to increase as
well. Sufficiently far from the open field line region the turbulent
transport can dominate these opposing contributions. Impurities tend
to reduce the turbulence level by dilution
\cite{pusztai11isotope,porkolab12,pusztai13}. If the non-diffusive
turbulent momentum transport is to be increased in spite of dilution
effects, stronger deviations of the non-fluctuating distribution from
a Maxwellian \cite{lee14} and stronger profile variations
\cite{waltz11} may be necessary; which could require a steepening of
the pedestal to reach a new steady state. Whether this is indeed the
case and, if so, its role in the observed confinement improvement in
impurity seeded discharges, remains an open question and should be the
basis of future investigation.


\acknowledgments The authors are grateful for J.~Omotani,
T.~F\"{u}l\"{o}p and S.~Newton for fruitful discussions and
instructive comments on the paper.  IP and SB were supported by the
International Career Grant of Vetenskapsr{\aa}det
(Dnr.~330-2014-6313), and ML was supported by the U.S. DoE under award
numbers DEFG0293ER54197 and DEFC0208ER54964.  The simulations used
computational resources of Hebbe at C3SE (project nr. C3SE2016-1-10
and SNIC2016-1-161).


\appendix
\section{\label{sec:profilesapp} The construction of model pedestal profiles}

Here, we describe a method to construct appropriate densities given
$T_a$, $\eta_z$, $\eta_i$ profiles satisfying \eqref{eq:ordering}, and
an arbitrary $n_i$. The orderings for the electrons are not a concern,
since $\rho_{pe}$ is much smaller than the radial scale length of any
profile in an experiment.

The density profiles and potential discussed in this section are the
inputs to \perfect{} and are thus, strictly speaking, only the zeroth
order contributions to these quantities, assumed to be flux functions.

Once we have specified both $\eta_i$ and $n_i$ (to be detailed shortly), the
relation $\eta_a = n_a e^{e_a \Phi/T_a}$ gives the potential
\begin{equation}
  \Phi_0 = \frac{T_i}{e_i} \log{\left(\frac{\eta_i}{n_i}\right)}.
\label{eq:generatephi} 
\end{equation}
A strong electric field from $\Phi_0$ compensates for the potentially
large ion pressure drop to make the variation of $\eta_i$ small (the
situation of ``electrostatic ion confinement,'' which is borne out in
experimental results \cite{viezzer13Thesis,mcdermott09}). In the radial force
balance the ion pressure drop could in principle be balanced by a
sonic ion flow, however we consider sub-sonic ion flows. Note that by
allowing sonic ion flows, the relative flow speed of two ion species
would in general also be sonic, which would severely complicate the
treatment of collisions. Such a scenario is currently not supported
by \perfect{}.  We emphasize that $\Phi_0$ and $n_a$ are fundamental
inputs to the code, thus the relation \eqref{eq:generatephi} is not an
attempt to self-consistently calculate the potential.

Then, given $\Phi_0$ from (\ref{eq:generatephi}) and an $\eta_z$, we
obtain the impurity density 
\begin{equation}
  n_z = \eta_z \left(\frac{n_i}{\eta_i}\right)^{\frac{e_z T_i}{e_i T_z }}.
\label{eq:nzprofile} 
\end{equation}
Note that (\ref{eq:nzprofile}) leads to impurity density profiles with
a typical logarithmic density gradient in the pedestal $Z_z$ times
larger than that of the main ions; this is unavoidable if
\eqref{eq:ordering} is to be satisfied for all ion species.  Finally,
the electron density is obtained by demanding quasi-neutrality
\begin{equation}
  n_e = Z_i n_i + Z_z n_z.
\end{equation}
The resulting $\eta_e$ satisfies \eqref{eq:ordering} by virtue of the
electron gyroradius being small.

Although we have constraints and relations between profiles, we still
have a large degree of freedom in specifying them. Here we discuss
some specific choices we made for the profile set used as a
baseline for the simulations.

Since we use the local solution for the boundary condition, we also
need to make sure that the assumptions of the local theory are
satisfied at the boundary. This means that the sharp density or
potential variations should be limited to the middle of the radial
domain, sufficiently far from the boundaries. For weak electric fields
the local and global results should agree, so we choose the $\eta_i$
profile so that the potential calculated from \eqref{eq:generatephi}
is completely flat at the boundaries. We do this by letting
\begin{equation}
\eta_i = n_i
\label{eq:lefteta}
\end{equation}
in the vicinity of the inner boundary, namely in the core up to the
pedestal top where $n_i$ (and so $\eta_i$) should be slowly varying.
Thus $\Phi_0=0$ in this region, independent of the $T_i$ profile. To
achieve a flat potential in the vicinity of the outer boundary, we let
\begin{equation}
\eta_i = n_i \exp{\left(C/T_i\right)},
\label{eq:righteta}
\end{equation}
in that region, where $C$ is a constant which fixes the value of
$e\Phi_0$. To remove the ambiguity of $\eta_i$ far from the boundaries
in a way that makes it a simple and smooth function, we linearly
extrapolate $\eta_i$ given by \eqref{eq:lefteta} from the core region
up to the bottom of the pedestal (where the sharp feature in $n_i$
ends) and match it to the other expression \eqref{eq:righteta},
choosing $C=e\Phi_0$ from \eqref{eq:generatephi} at the matching
point.

The $\eta_z$ profile is chosen to be a linear function of $\psi$ over
the whole domain, with a logarithmic gradient matching that of $n_i$ at
the left boundary.

Since the global $\delta f$ ordering does not allow an ion temperature
pedestal, we consider $T_i=T_z$ profiles with a gradient across the
pedestal (and further out) equal to that of the electron temperature
gradient in the core. Due to the density drop in the pedestal and the
decreasing temperature profile, the ion heat flux is bound to be
vastly different at the two boundaries, requiring large heat sources
in the domain, except if a variation in $\d T_i/\d\psi$ balances
them. This can be avoided by setting a proxy for the heat fluxes,
$\propto n_i T_i^{3/2}
\d T/\d \psi$, to be equal at the boundaries. Thus, we artificially
reduce $\d T_i/\d\psi$ in the core to remove the need for sources close
to the boundaries.

Experimental density and temperature profiles can drop orders of
magnitude across the pedestal, and a significant part of these
variations occur in the open field line region.  As an additional
consideration, if we let the bulk densities and temperatures drop
across the simulation domain as much as in a real pedestal, it would
lead to difficulties related to large logarithmic gradients, and huge
changes in collision frequency. Since the code does not capture the
physics in the open field line region (where the pedestal foot would
be in an experiment), the region outside the middle of the pedestal
does not carry too much physical relevance and can be considered as a
numerical buffer zone. To avoid the above mentioned complications we
reduce the gradients starting from a point where crossing the
separatrix would be expected in an experiment. We thus arbitrarily
pick this reduced $\d T_e/\d \psi$ to be $0.05$ of its pedestal value,
with $\d n_i/\d\psi$ equal to its core value.

Motivated by typical JET discharges (Figure 16 of
\citenum{0029-5515-55-11-113031}), for our baseline case we choose
the $T_e$ and $n_i$ pedestal widths to be $\sim \unit[3]{cm}$, with
the values $T_{e}=\unit[0.9]{keV}$ and $n_{e}=\unit[4\cdot
10^{19}]{m^{-3}}$ at the pedestal top. These correspond to typical
logarithmic gradients of about $-\d(\ln n_e)/\d(r/a) =
32.55$ and $-\d(\ln T_e)/\d(r/a) = 32.56$, where $r$ is the minor radius
defined as half of the width of the flux surface at the elevation
of its centroid, and $r=a$ at the last closed flux surface.  The $T_e$
and $n_i$ profiles are generated by Bezier curve interpolation between
three regions with linear profile variation: a core, a pedestal, and
an outer ``buffer'' region.  (Using this type of smoothing ensures
that the gradients transition smoothly and monotonically.)

\begin{figure}
\includegraphics[width=1.0\columnwidth]{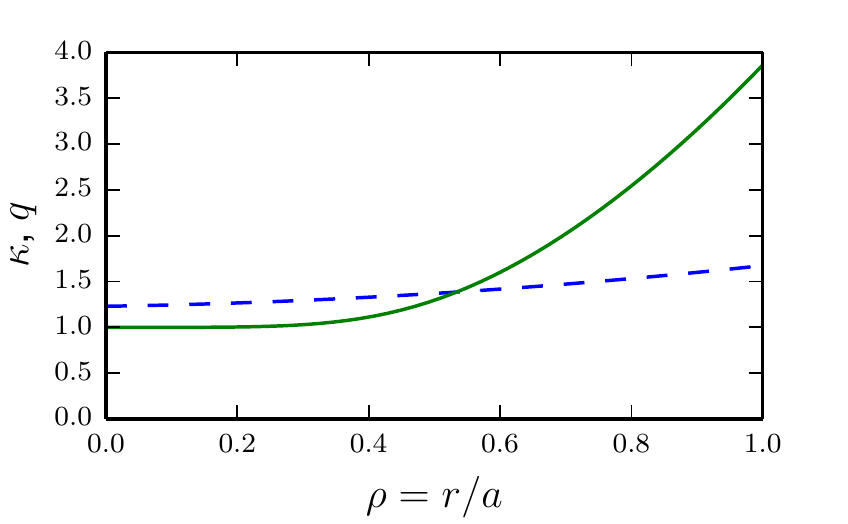}
\caption{\label{fig:qkappa} $q$ (solid curve) and $\kappa$ (dashed)
  profiles used to calculate $\psi(r)$.}
\end{figure}

To transform the profiles from $r$ to $\psi_N$ space, $\psi(r)$ is
needed.  This we obtain from $q(r) = \d\chi/\d\psi$ where $2\pi \chi$
is the toroidal magnetic flux which we calculate assuming simple
elongated flux surfaces.  Since we do not intend to model a specific
experiment, we take model profiles for the safety factor $q$ and the
elongation $\kappa$, shown in \autoref{fig:qkappa}.  The $q_{95}$ was
chosen as $3.5$.  For the on-axis toroidal magnetic fields we take
$B_t=\unit[2.9]{T}$ and neglect $\epsilon^2$ corrections together with
higher order shaping effects to get $\d r/\d\psi_N$.  The resulting
$\d r/\d\psi_N|_{\rm LCFS} = 0.57$ is taken to be constant across the
entire pedestal.

\begin{figure}
\includegraphics[width=1.0\columnwidth]{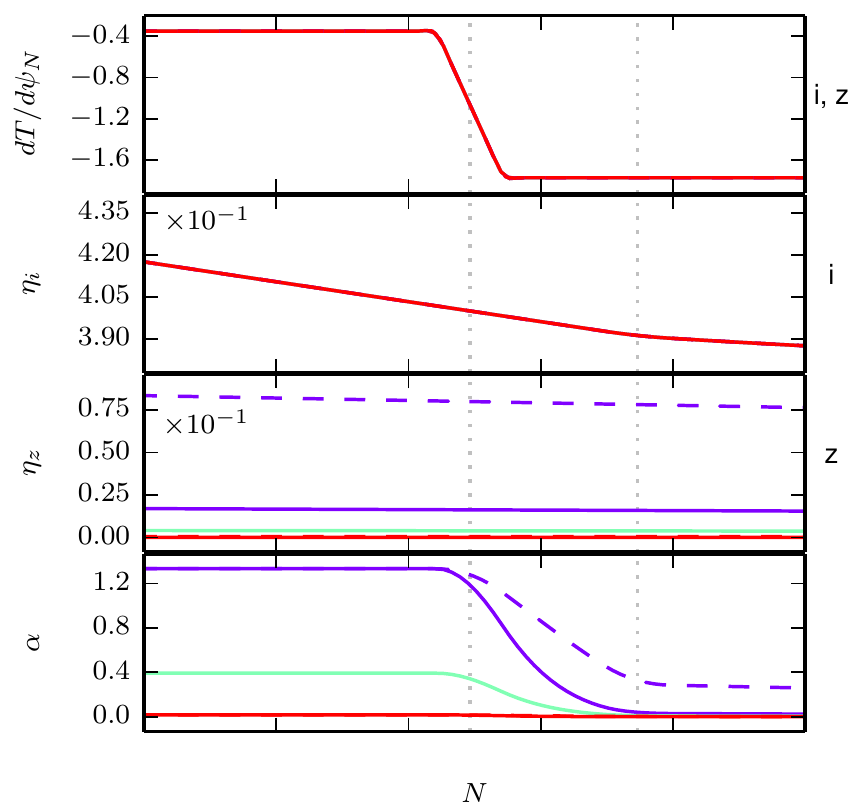}
\caption{\label{fig:extraInput} Ion temperature gradient [$\rm
      keV$], $\eta$ [$10^{20}\rm m^{-3}$], and impurity strength
    profiles for the different simulations presented in this paper:
    higher impurity density (violet), baseline (cyan), trace impurity
    (red). Dashed lines correspond to the beryllium case of
    Fig.~\ref{fig:Z4conc_momsum}. Species is indicated
    to the right of the panels.}
\end{figure}

The ion temperature gradient, the ion and impurity $\eta$ and the
impurity strength profiles are shown in \autoref{fig:extraInput}, for
the different impurity concentration simulations presented in this
paper.

\section{\label{sec:resolapp} Numerical resolution}

The simulations used $N_\psi=204$ radial grid points and cover a
domain of $\psi_N=0.85-1.1$ (the whole $\psi_N$ domain is not shown in
the figures). The number of poloidal grid points is $N_\theta=75$. The
number of expansion polynomials in the pitch angle cosine $\xi=v_\|/v$
is $N_\xi=24$ for the distribution function and $N_\xi^{\rm RP}=4$ for
the Rosenbluth potentials (RP). The number of speed grid points is
$N_x=8$ and $N_x^{\rm RP}=150$ for the RP.

\begin{figure}
    \includegraphics{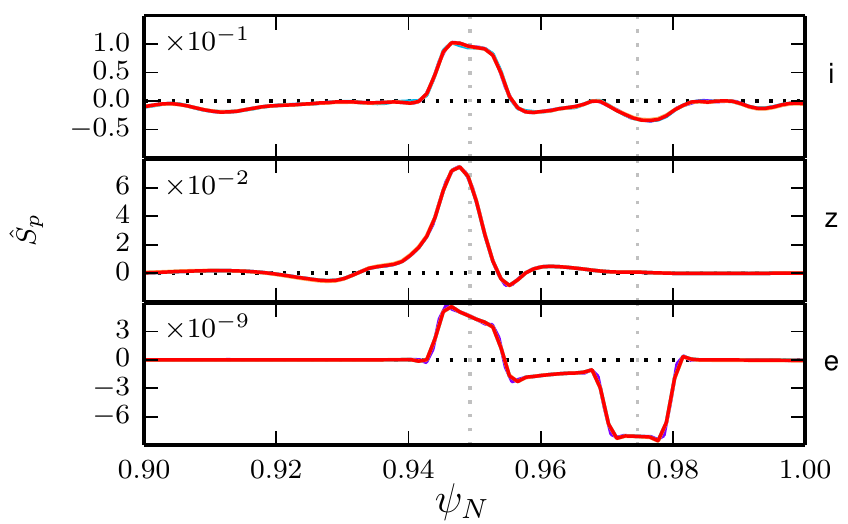}
\put(-25,124){\large a}
\put(-25,84){\large b}
\put(-25,52){\large c}
\caption{\label{num_psource} The particle sources from a baseline
  simulation (red), plotted alongside simulations with $N_\xi=30$
  (cyan), $N_\theta=90$ (blue), $N_\psi=255$ (violet), radial
    domain size $\psi_N \in [0.825,1.125]$ (yellow). Species
    is indicated to the right of the panels.} 
\end{figure}

To demonstrate the degree of convergence, in \autoref{num_psource} we
present the particle sources for the baseline simulation for the above
mentioned resolution (red curve), together with four other cases,
where we increase $N_\psi$ to $255$ (violet), $N_\theta$ to
$N_\theta=90$ (blue), $N_\xi$ to $30$ (cyan), and the radial domain
size to $\psi_N \in [0.825,1.125]$ (yellow), with all other parameters
kept fixed. These four are the resolution parameters to which the
accuracy of the solution is most sensitive. The results are almost
identical except for slight differences near sharp features.

To quantify the error, we define ${\rm err}(X_1)= \left(\int_{0.9}^{1}
|X_2-X_1| d\psi_N \right)/\left(\int_{0.9}^{1} |X_1| d\psi_N \right).$
Taking the $N_\psi=255$ simulation as the reference $X_2$, we find
that the quantity with the highest error is the electron particle
source, with ${\rm err}(\hat{S}_{pe})=5.6\%$.  For other quantities,
such as particle flows and radial fluxes, we observe errors below
$0.5\%$, thus \autoref{num_psource} presents the most stringent test
for numerical convergence. The same convergence test was performed for
all the simulations, with resulting errors (including those of the
sources) all being below $6\%$.

\bibliography{plasma-bib.bib} 
\end{document}